\documentclass[aps,prl,twocolumn,groupedaddress,showpacs,superscriptaddress]{revtex4-1}
\usepackage{wasysym}
\usepackage{hyperref}
\usepackage{graphicx}
\usepackage{color}

\begin{document}
\title{The Mamoudou Gassama Affair}

\author{V.G.~Rousseau}
\affiliation{Physics Department, Loyola University New Orleans, 6363 Saint Charles Ave., New Orleans, Louisiana 70118, USA}

\begin{abstract}
On May $26^{\textrm{\tiny th}}$, 2018, it was reported in the French news that a four-year-old child fell from the $5^{\textrm{\tiny th}}$ floor's balcony
of a building in Paris (France), and was able to catch the railing of the $4^{\textrm{\tiny th}}$ floor's balcony. An immigrant, Mamoudou Gassama, who was passing by
decided to climb the building and rescued the child. Subsequently, Mamoudou was congratulated by French president, Emmanuel Macron, and was naturalized.
In this paper, by using kinematic equations and Newton's laws, it is shown that it is actually impossible for a four-year-old child (and probably for an adult too) to fall from a balcony and catch
the railing of the lower balcony. This suggests that the rescue of the child could have been staged.

The author's interview by famous French Journalist Andr\'e Bercoff was broadcast on \textit{Sud Radio} on June $4^{\textrm{\tiny th}}$, 2019,  and is available at:\\
\href{https://www.youtube.com/embed/AHkUhrtM2B0}{https://www.youtube.com/embed/AHkUhrtM2B0}
\end{abstract}

%\pacs{89.90.+n}
\maketitle

\section*{Introduction}
It was about 8:00 pm on Saturday, May $26^{\textrm{\tiny th}}$, 2018, when firemen were alerted by pedestrians that a child was hanging on the railing of
the $4^{\textrm{\tiny th}}$ floor of a building in Paris (France)\cite{LCI,Parisien,FranceSoir1,CNews,China}. When they arrived on the scene, the child had been rescued by Mamoudou Gassama, a Malian immigrant.
The scene was recorded\cite{LCI}, showing Mamoudou climbing the four stories in about 30 seconds, grabbing the kid's arm, lifting him over the railing, and
putting him in safety. Subsequently, Mamoudou was congratulated by president Emmanuel Macron, who proposed to engage right away a procedure of naturalization,
which Mamoudou accepted\cite{FranceSoir2}.

Following the online publication of the present study, the author was interviewed by famous French journalist Andr\'e Bercoff, and the interview
was broadcast on \textit{Sud Radio} on June $4^{\textrm{\tiny th}}$, 2019\cite{Bercoff}.

One question that arises is how the four-year-old child ended up hanging on the $4^{\textrm{\tiny th}}$ balcony's railing, where supposedly nobody else was
present, and where the windows were locked from the inside. It was reported that the child didn't talk but indicated with his finger that
he fell from above\cite{LCI,Parisien,FranceSoir1,CNews,China}, presumably the $5^{\textrm{\tiny th}}$ floor, although there is no testimony of anybody having seen him falling. However, the
concierge of the building later declared that the $5^{\textrm{\tiny th}}$ floor is uninhabited, which suggests that the child fell from the $6^{\textrm{\tiny th}}$ floor, where he lives\cite{China}.

In this paper, it is shown by using kinematic equations and Newton's laws that the above scenario is impossible. It is important here
to point that we don't claim that the rescue of the child by Mamoudou is staged, we only consider it as a possibility. The only claim that we make is that, as opposed to what was reported in the news, the 
child didn't fall from one or more stories. We don't make any hypotheses on whether the child was put on the $4^{\textrm{\tiny th}}$ balcony's railing by irresponsible parents in order to provide
Mamoudou with an opportunity to accomplish his exploit, or if Mamoudou was totally unaware of the reasons why the child was hanging up there. We also don't
even comment on the fact that the child didn't lose his flip-flops during the reported fall, and leave it to the reader's consideration. Indeed, Mamoudou
declared that once he put the child in safety, he noticed that he was wearing Spiderman flip-flops,\cite{20minutes} a funny coincidence for the so-called ``French Spiderman".\\

\section*{Model}
\begin{figure}[h]
\centerline{\includegraphics[width=0.5\textwidth]{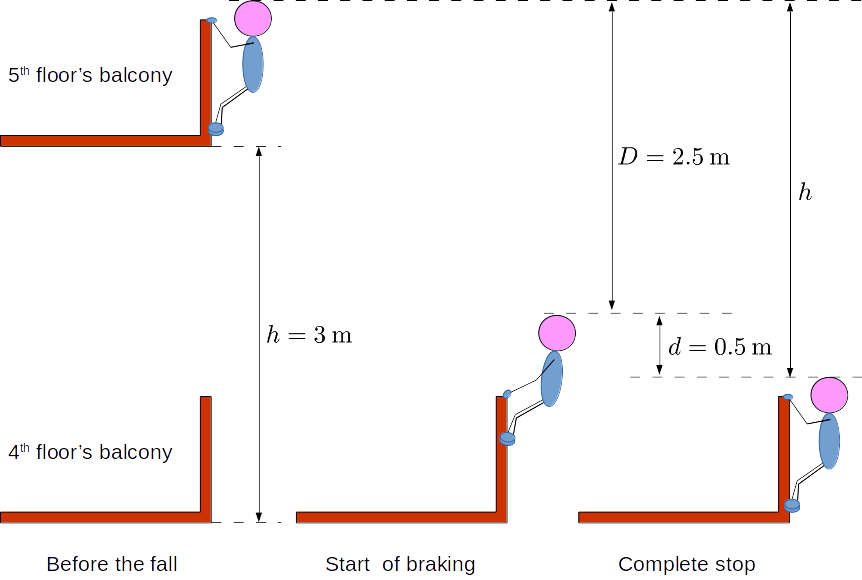}}
\caption{(Color online) Ideal situation where the child falls from the height of a single story, $h=3.00\:\rm m$, and slows down and comes to a complete stop over a distance of $d=50\:\rm cm$.}
\label{Mamoudou}
\end{figure}

In the following, a calculation that largely underestimates the force that the child would have had to produce in order to stop his fall is presented, so that it
clearly shows that such an exploit is impossible. Although the child supposedly fell from two stories, we assume that he fell from only one, Fig.~\ref{Mamoudou}, and that the distance between two consecutive balcony railings is
$h=3.00\:\rm m$ (standard distance). In order to make sure that the calculated average force is underestimated, the ideal case where the child manages to slow
down and come to a stop over the largest possible distance is considered. Typically, the braking distance is equal to about the child's arm length. Considering that, on
average, the arm length of a four-year-old child is $11.0\:\rm in$, this corresponds to a distance of $27.9\:\rm cm$. However, in the following we purposely
overestimate this distance and take it to be $d=50.0\:\rm cm$, in order to make sure that the calculated average force is underestimated.

As is shown below, the velocity that the child reaches before catching the railing is much smaller than the terminal velocity of a skydiver. This justifies
that air resistance can be neglected. Also, during the fall, since there is no horizontal force pushing the child against the railing and the balcony's concrete, the friction force due to his nails,
clothes, and flip-flops rubbing against the railing and the balcony's concrete vanishes quickly as soon as he starts to fall. Indeed, this kinetic
friction force is proportional to the normal (horizontal) force that the railing and balcony's concrete exert on the child. If this normal force is initially present, it is
the only horizontal force acting on the child, who is therefore accelerated away from the building. Thus the force that he is able to apply on it (equal in magnitude to the normal reaction force)
quickly vanishes. This friction force can therefore
be neglected as well. As a result, during the fall, only gravity is at play so that the child is in a free fall.

\section*{Kinematic equations}
The child is in a free fall over a distance $D$ equal to the distance between the balconies, minus the braking distance, $D=h-d=2.50\:\rm m$. During this 
free fall, the child's acceleration is constant, equal to $g=9.80\:\rm m/s^2$. Using the well known kinematic equations for constant acceleration,
\begin{equation}
y=y_o+v_ot+\frac12 at^2\quad\quad\quad v=v_o+at,
\end{equation}
where $y,y_o,a,t,v,v_o$ are respectively the final height, initial height, acceleration due to gravity, time, final velocity, and initial velocity,
and substituting $v_o=0$ (child initially at rest), we can eliminate the time and express the velocity
as a function of the free-fall distance $D=|y-y_o|$:
\begin{equation}
\label {Math1} v=\sqrt{2a|y-y_o|}
\end{equation}
Substituting $|y-y_o|=2.50\:\rm m$, and $a=g=9.80\:\rm m/s^2$, we get:
\begin{equation}
v=7.00\:\rm m/s=25.2\:\rm km/h
\end{equation}
This is the child's velocity right before he catches the railing. Note that, as already announced, this velocity is much smaller than the terminal velocity
of a skydiver (about 200 km/h), which justifies neglecting air resistance. At this point, one could think that the challenge for the child is to catch the railing. But
even more challenging is for him to slow down and come to a complete stop over the distance $d$. Indeed, let us calculate the average acceleration (deceleration)
needed by solving Eq.~\ref{Math1} for $a$:
\begin{equation}
a=\frac{v^2}{2|y-y_o|}
\end{equation}
Substituting $v=7.00\:\rm m/s$ and the braking distance $|y-y_o|=0.500\:\rm m$, the average acceleration needed to come to a stop is:
\begin{equation}
a=49.0\:\rm m/s^2
\end{equation}
This corresponds exactly to an average acceleration of $5g$.\\

\section*{Newton's $\bf 2^{\textrm{\tiny nd}}$ and $\bf 3^{\textrm{\tiny rd}}$ laws}
\begin{figure}[h]
\centerline{\includegraphics[width=0.35\textwidth]{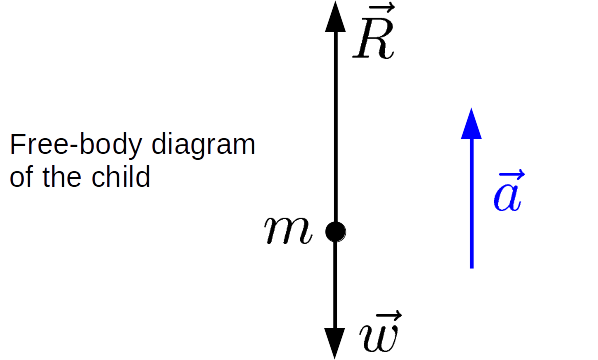}}
\caption{(Color online) Free-body diagram of the child, showing his weight $\vec w$ and the railing's reaction force $\vec R$. The acceleration $\vec a$ is also shown (the 
child is slowing down with a downward velocity, thus the net acceleration is upward).}
\label{Mamoudou2}
\end{figure}

Let us denote by $m$ the mass of the child.
During the braking, two forces are acting on him: His weight $\vec w$ with magnitude $w=mg$, and the railing's reaction force $\vec R$, as shown on the free-body diagram,
Fig.~\ref{Mamoudou2}.

According to Newton's $2^{\textrm{\tiny nd}}$ law, the net force $\vec F_\textrm{\tiny net}$ acting on the child is:
\begin{equation}
\vec F_\textrm{\tiny net}=\vec R+\vec w=m\vec a
\end{equation}
Solving for $\vec R$, we get:
\begin{equation}
\vec R=m\vec a-\vec w
\end{equation}
In term of magnitudes, this becomes:
\begin{equation}
R=m(a+g)
\end{equation}
Substituting $a=49.0\:\rm m/s^2$, $g=9.80\:\rm m/s^2$, and assuming an average mass of $20.0\:\rm kg$ for a four-year-old child, the average force exerted by the
railing onto the child has a magnitude of:
\begin{equation}
R=1176\:\rm N
\end{equation}
According to Newton's $3^{\textrm{\tiny rd}}$ law, this force is equal in magnitude and opposite in direction to the force exerted by the child against the railing.
This force is precisely equal to {\bf six times his own weight} (this holds true for any mass $m$). In other words, this is the force necessary to lift a mass of 120 kg. It is hard to believe
that a four-year-old child could be able to accomplish such an exploit.
Also, note that this is just the average force during the braking. The instantaneous force could easily be an order of magnitude greater. In addition,
this average force is way underestimated, since we have purposely overestimated the braking distance $d$, and considered a fall from the height of a single story,
whereas the child supposedly fell from two. As a result, it is clear that the reported scenario is impossible.

Another way to realize that the scenario is impossible is to imagine that we ask a four-year-old child to catch a mass of $20.0\:\rm kg$ that is dropped
$3.00\:\rm m$ from above...

\section*{Critics of the model}
A common critic of the model is that it does not take into account frictional forces due to the child trying to grab anything on his way during the fall. As
explained in section ``model", the kinetic friction force quickly vanishes as soon as the child starts to fall. Also, it was reported by the neighbor on the
$4^{\textrm{\tiny th}}$ floor (who didn't attempt anything to rescue the child) that he noticed that the child had a torn toe nail\cite{LCI}. Assuming that this information is correct, the force necessary to tear the nail must be
compared to the force of $1176\:\rm N$ necessary to stop the fall. Common sense clearly allows us to conclude that these two forces cannot compete.

\begin{figure}[h]
\centerline{\includegraphics[width=0.5\textwidth]{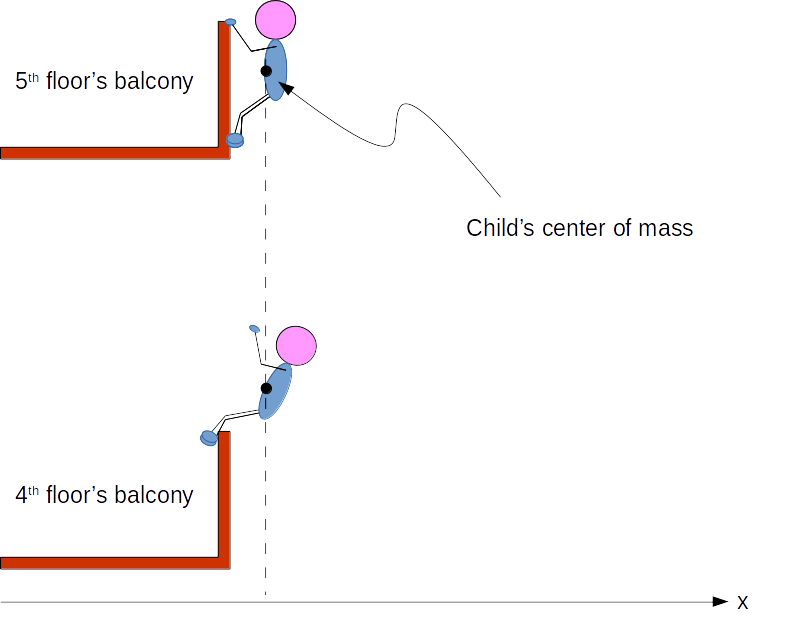}}
\caption{(Color online) Because horizontal momentum is conserved, the feet or legs hitting first the railing would make the child fall backward, eliminating
any possibility of catching the railing with his hands.}
\label{Mamoudou3}
\end{figure}

Another common critic is that the child could have slowed down his fall with his feet or legs hitting the railing first. This doesn't make any sense either,
since conservation of horizontal momentum guarantees that the child's center of mass cannot move towards the railing. If during the fall the feet or legs of the child move
toward the building, then his upper body must necessarily move away from it, as illustrated in Fig.~\ref{Mamoudou3}. He would therefore fall backward,
without any chance of catching the railing with his hands. In addition, such a collision with the railing would clearly have led to injuries, which
have not been reported.

\section*{Conclusion}
It was reported in worldwide news that the child rescued by Mamoudou Gassama fell from one or more stories\cite{LCI,Parisien,FranceSoir1,CNews,China}.
However, in this paper, it is shown that this scenario is impossible.
This raises the question about determining how the child actually ended up hanging
on the railing of the $4^{\textrm{\tiny th}}$ balcony, where supposedly nobody else was present, while he lives on the $6^{\textrm{\tiny th}}$ floor, where it was reported by his father
that he was left alone. Given these facts, it
is hard to avoid the idea that the rescue of the child could have been staged. It should also be noted that one could have arrived to the same conclusion (namely the scenario being impossible)
without any calculations, just by using common sense. More than one month after the facts, it is very surprising that nobody has made a public claim that
the reported scenario is impossible. It is even more surprising that, presumably, president Emmanuel Macron wasn't advised about the 
glitches of the case, and decided to proceed right away with the naturalization of Mamoudou\cite{FranceSoir2}.

%\begin{acknowledgments}
%None! This work hasn't received any support. On the contrary, I made several attempts to publish my calculations and conclusions on several forums, but my posts
%were systematically deleted by the webmasters, although no insults or hateful language was used.
%s\end{acknowledgments}

\end{document}